\documentclass[11pt,showpacs,superscriptaddress,amsmath,amssymb,nofootinbib]{revtex4}

\usepackage{graphicx}
\usepackage{dcolumn}
\usepackage{bm}
\usepackage{amssymb}
\usepackage{amsmath}
\usepackage{epsfig}
\usepackage{color}
\usepackage{slashed}
\usepackage{float}
\usepackage{longtable}
\usepackage{braket}
\usepackage{ulem}
\def\cf{c_\phi}
\def\sf{s_\phi}

\def\ol{\overline}

\newcolumntype{I}{!{\vrule width 1.3pt}}
\begin{document}

\centerline{\large\bf A global analysis of two-body $D \to V\!P$ decays}
\centerline{\large\bf  within the framework of flavor symmetry}

\author{Hai-Yang Cheng}
\email{phcheng@phys.sinica.edu.tw}
\affiliation{Institute of Physics, Academia Sinica, Taipei, Taiwan 11529, Republic of China}
\author{Cheng-Wei Chiang}
\email{chengwei@ncu.edu.tw}
\affiliation{Institute of Physics, Academia Sinica, Taipei, Taiwan 11529, Republic of China}
\affiliation{Department of Physics and Center for Mathematics and Theoretical Physics,
National Central University, Chungli, Taiwan 32001, Republic of China}
\affiliation{Physics Division, National Center for Theoretical Sciences, Hsinchu, Taiwan 30013, Republic of China}
\author{An-Li Kuo}
\email{101222028@cc.ncu.edu.tw}
\affiliation{Department of Physics and Center for Mathematics and Theoretical Physics,
National Central University, Chungli, Taiwan 32001, Republic of China}

\begin{abstract}
Two-body charmed meson decays $D\to VP$ are studied within the framework of the diagrammatic approach. Under flavor SU(3) symmetry, all the flavor amplitude sizes and their associated strong phases are extracted by performing a $\chi^2$ fit. Thanks to the recent measurement of $D_s^+\to\pi^+\rho^0$, the magnitudes and the strong phases of the $W$-annihilation amplitudes $A_{P,V}$ have been extracted for the first time. As a consequence, the branching fractions of all the $D\to VP$ decays are predicted, especially those modes that could not be predicted previously due to the unknown $A_{P,V}$. Our working assumption, the flavor SU(3) symmetry, is tested by comparing our predictions with experiment for the singly and doubly Cabibbo-suppressed decay modes based on the flavor amplitudes extracted from the Cabibbo-favored decays using the current data. The predictions for the doubly Cabibbo-suppressed channels are in good agreement with the data, while those for the singly Cabibbo-suppressed decay modes are seen to have flavor SU(3) symmetry breaking effects. We find that the inclusion of SU(3) symmetry breaking in color-allowed and color-suppressed tree amplitudes is needed in general in order to have a better agreement with experiment. Nevertheless, the exact flavor SU(3)-symmetric approach alone is adequate to provide an overall explanation for the current data.
\end{abstract}


\maketitle
\newpage

\section{Introduction \label{sec:intro}}

Recently, there were some new measurements of the $D$ meson decaying into a pseudoscalar meson $P$ and a vector meson $V$, such as the branching fractions of $D^+\to\pi^+\omega, D^0\to\pi^0\omega$ and several doubly Cabibbo-suppressed decay modes.  Such information enables us to test how well flavor SU(3) symmetry holds in the system.  The $D\to VP$ decays had been studied in the diagrammatic approach~\cite{Chau:1987tk, Cheng:2010ry, Bhattacharya:2008ke} as well as in the perturbative approach~\cite{Buccella:1994nf, Li:2013xsa, Zou:2013ez}.  Under the assumption of SU(3)$_F$ flavor symmetry, quark diagrams of the same topology, including the associated strong phases, are identical to one another, modulo the obvious different Cabibbo-Kobayashi-Maskawa (CKM) matrix elements.  We will adopt this symmetry as our working assumption in this paper.  In particular, we will extract information of the flavor diagrams through a $\chi^2$ fit to the Cabibbo-favored decay modes.

In our previous work~\cite{Cheng:2010ry}, we had shown that the $W$-annihilation amplitudes $A_{P,V}$ could not be completely determined based on the data available at that time.  Consequently, many of the $D^+$ and $D_s^+$ decays that involve the $A_{P,V}$ amplitudes could not be predicted within the framework of SU(3)$_F$ symmetry.

In this work, we not only update the analysis based on the latest data, but in particular extract information (the magnitudes and associated strong phases) of the $A_{P,V}$ amplitudes for the first time, thanks to the recent measurement of $D_s^+\to\pi^+\rho^0$ branching fraction.  As a result, we are able to make predictions for all the decay rates without additional assumptions.  More explicitly, we determine all tree-level flavor amplitudes from the Cabibbo-favored decay modes through a $\chi^2$ fit.  Based on several comparable fit solutions, we then make predictions for the singly and doubly Cabibbo-suppressed decay modes using the SU(3)$_F$ symmetry.  We observe again flavor SU(3) symmetry breaking effects in certain singly Cabibbo-suppressed modes.  We then study whether such effects can be accounted for by considering factorization for color-allowed and color-suppressed tree amplitudes $T_{P,V}$ and $C_{P,V}$ and including ratios of decay constants, and form factors among modes of different Cabibbo factors.  The result is also compared with the effective Wilson coefficients $a_{1,2}$ calculated by perturbation.

This paper is organized as follows.  In Section~\ref{sec:data}, we present the current experimental data of all the $D \to VP$ decay channels.  We discuss how to extract those observables that we are interested in from experiment.  In Section~\ref{sec:formalism}, we review flavor amplitude decomposition of all the decay modes and the convention used in this work, based on the SU(3)$_F$ symmetry.  In Section~\ref{sec:fitting}, we perform a $\chi^2$ fit to the data of the Cabibbo-favored modes, thereby extracting the central values and 1-$\sigma$ ranges of the magnitude and strong phase for each flavor amplitude.  Solutions of similar fit quality are all presented.  Based on these solutions, we make predictions for all the $D\to VP$ branching fractions in Section~\ref{sec:predictions}.  In Section~\ref{sec:SU(3)breaking}, we discuss possible SU(3)$_F$ symmetry breaking effects from the differences in decay constants, and form factors for color-allowed and color-suppressed tree amplitudes.  Finally, the conclusions are given in Section~\ref{sec:conclusions}.

\section{Experimental Data \label{sec:data}}

Before presenting the data of all the $D\to VP$ decay modes, we note that some of them are extracted from three-body decays through a vector-meson resonance; that is, $D\to P_1P_2P_3$ through $V\to P_1P_2$ with $V$ being $K^*$ or $\phi$.  Under the narrow width approximation, ${\cal B}(D\to P_1P_2P_3)={\cal B}(D\to VP_3){\cal B}(V\to P_1P_2)$.  The branching fractions of such modes are given in Table~\ref{extract}.
To obtain the experimental branching fractions of the associated $D\to VP$ decays, we make use of the following branching fractions
\begin{equation}
\begin{split}
{\cal B}(K^{*-}\to K^-\,\pi^0) &=\frac{1}{3} ~,\\
{\cal B}(K^{*-}\to K_S\,\pi^-) &=\frac{1}{3} ~,\\
{\cal B}(\ol K^{*0}\to K^-\,\pi^+) &=\frac{2}{3} ~,\\
{\cal B}(\ol K^{*0}\to K_S\,\pi^0) &=\frac{1}{6} ~,\\
{\cal B}(\phi\to K^+\,K^-) &=(48.9\pm0.5)\% ~.
\end{split}
\label{extract}
\end{equation}
Under the assumption that ${\cal B}(K^*\to K\pi)=100\%$, the first four relations in Eq.~(\ref{extract}) follow from isospin symmetry.  Note that a factor of 2 should be multiplied when going from the branching fractions of modes with $K_S$ to those of modes with $K^0$ or $\ol{K}^0$.  For those channels whose vector mesons can decay into more than one channel, we take their weighted averages.  Along with the other modes, all available averaged experimental branching fractions are listed in Table~\ref{VPCF}, Table~\ref{VPSCS} and Table~\ref{VPDCS} for Cabibbo-favored, singly Cabibbo-suppressed and doubly Cabibbo-suppressed decay modes, respectively.  Unless specified, meson masses, lifetime and all the branching fraction data are taken from the Particle Data Group (PDG)~\cite{pdg}.  Any asymmetric uncertainties are averaged for simplicity.

\begin{table}[tp!]
\footnotesize{
\caption{Branching fractions of some $D\to P_1P_2P_3$ decays through a vector meson resonance.}

\vspace{6pt}
\begin{ruledtabular}
\begin{tabular}{l l}
${\cal B}(D\to VP){\cal B}(V\to PP)$ & ${\cal B}(D\to VP)$
\\
\multicolumn{2}{c}{$(\%)$}
 \\
 \hline
 $\begin{array}{c}
{\cal B}(D^0\to K^{*-}\,\pi^+){\cal B}(K^{*-}\to K_S\,\pi^-)=1.68^{+0.15}_{-0.18}\\
{\cal B}(D^0\to K^{*-}\,\pi^+){\cal B}(K^{*-}\to K^-\,\pi^0)=2.28^{+0.40}_{-0.23}
 \end{array}\Bigg\}$
 &${\cal B}(D^0\to K^{*-} \pi^+)=5.43\pm0.44$  \\

  $\begin{array}{c}
{\cal B}(D^0\to \ol{K}^{*0}\,\pi^0){\cal B}(\ol K^{*0}\to K^-\,\pi^+)=1.93\pm0.26\\
{\cal B}(D^0\to \ol{K}^{*0}\,\pi^0){\cal B}(\ol K^{*0}\to K_S\,\pi^0)=0.79\pm0.07
 \end{array}\Bigg\}$
 &${\cal B}(D^0\to \ol{K}^{*0} \pi^0)=3.75\pm0.29$ \\

  $\begin{array}{c}
{\cal B}(D^+\to \ol{K}^{*0}\,\pi^+){\cal B}(\ol K^{*0}\to K^-\,\pi^+)=1.05\pm0.12\\
{\cal B}(D^+\to \ol{K}^{*0}\,\pi^+){\cal B}(\ol K^{*0}\to K_S\,\pi^0)=0.259\pm0.031
 \end{array}\Bigg\}$
 & ${\cal B}(D^+\to\ol{K}^{*0} \pi^+)=1.57\pm0.13$\\

${\cal B}(D^0\to K_S\,\rho^0)=0.64^{+0.07}_{-0.08}$                                         &${\cal B}(D^0\to \ol{K}^0\,\rho^0)=1.28^{+0.14}_{-0.16}$              \\
${\cal B}(D^0\to \ol{K}^{*0}\,\eta){\cal B}(\ol K^{*0}\to K_S\,\pi^0)=0.16\pm0.05$ &${\cal B}(D^0\to \ol{K}^{*0}\,\eta)=0.96\pm0.30$              \\
${\cal B}(D^0\to K_S\,\omega)=1.11\pm0.06$                                                     &${\cal B}(D^0\to \ol{K}^0\,\omega)=2.22\pm0.12$              \\
${\cal B}(D^0\to K_S\phi){\cal B}(\phi\to K^+K^-)=0.207\pm0.016$                  & ${\cal B}(D^0\to\phi\,K^0)=0.847^{+0.066}_{-0.034}$\\

${\cal B}(D^+\to K_S\,\rho^+)=6.04^{+0.60}_{-0.34}$                                         &${\cal B}(D^+\to \ol{K}^0\,\rho^+)=12.08^{+1.20}_{-0.68}$              \\
${\cal B}(D_s^+\to\ol{K}^{*0}\,K^+){\cal B}(\ol K^{*0}\to K^-\,\pi^+)=2.61\pm0.09$ &${\cal B}(D_s^+\to\ol{K}^{*0}\,K^+)=3.92\pm0.14$  \\
\hline\hline
   \multicolumn{2}{c}{$(\times 10^{-3})$}
   \\
\hline
${\cal B}(D^0\to K^+\,K^{*-}){\cal B}(K^{*-}\to K^-\,\pi^0)=0.54\pm0.05$        & ${\cal B}(D^0\to K^+\,K^{*-})=1.62\pm0.15$ \\
${\cal B}(D^0\to K^-\,K^{*+}){\cal B}(K^{*+}\to K^+\,\pi^0)=1.50\pm0.10$      & ${\cal B}(D^0\to K^-\,K^{*+})=4.50\pm0.30$  \\
${\cal B}(D^0\to K_S\,\ol{K}^{*0}){\cal B}(\ol K^{*0}\to K^-\,\pi^+)<0.5$         & ${\cal B}(D^0\to K^0\,\ol{K}^{*0})<1.5$ \\
${\cal B}(D^0\to K_S\,K^{*0}){\cal B}(K^{*0}\to K^+\,\pi^-)<0.18$            &${\cal B}(D^0\to \ol{K}^0\,K^{*0})<0.54$\\
${\cal B}(D^0\to\pi^0\,\phi){\cal B}(\phi\to K^+\,K^-)=0.66\pm0.05$               &${\cal B}(D^0\to\pi^0\,\phi)=1.35\pm0.10$\\
${\cal B}(D^+\to\pi^+\,\phi){\cal B}(\phi\to K^+\,K^-)=2.77^{+0.09}_{-0.10}$                      &${\cal B}(D^+\to\pi^+\,\phi)=5.66^{+0.19}_{-0.21}$  \\
${\cal B}(D^+\to K^+\,\ol{K}^{*0}){\cal B}(\ol K^{*0}\to K^-\,\pi^+)=2.56^{+0.09}_{-0.15}$&${\cal B}(D^+\to K^+\,\ol{K}^{*0})=3.84^{+0.14}_{-0.23}$  \\
${\cal B}(D^+\to K_S\,K^{*+})=17\pm8$                                                                              &${\cal B}(D^+\to\ol{K}^0\,K^{*+})=34\pm16$  \\
${\cal B}(D_s^+\to\pi^+\,K^{*0}){\cal B}(K^{*0}\to K^+\,\pi^-)=1.42\pm0.24$ &${\cal B}(D_s^+\to\pi^+\,K^{*0})=2.13\pm0.36$  \\
${\cal B}(D_s^+\to K^+\,\phi){\cal B}(\phi\to K^+\,K^-)=0.089\pm0.020$       & ${\cal B}(D_s^+\to K^+\,\phi)=0.164\pm0.041$ \\
\hline\hline
\multicolumn{2}{c}{$(\times 10^{-4})$}
 \\
\hline
${\cal B}(D^0\to K^{*+}\,\pi^-){\cal B}(K^{*+}\to K_S\,\pi^+)=1.15^{+0.60}_{-0.34}$& ${\cal B}(D^0\to K^{*+}\,\pi^-)=3.45^{+1.80}_{-1.02}$\\
${\cal B}(D^+\to K^{*0}\,\pi^+){\cal B}(K^{*0}\to K^+\,\pi^-)=2.6\pm0.4$ &${\cal B}(D^+\to K^{*0}\,\pi^+)=3.9\pm0.6$  \\
${\cal B}(D_s^+\to K^{*0}\,K^+){\cal B}(K^{*0}\to K^+\,\pi^-)=0.60\pm0.34$& ${\cal B}(D_s^+\to K^{*0}\,K^+)=0.90\pm0.51$ \\
\end{tabular}
\label{VPCF}
\end{ruledtabular}
}
\end{table}

It has long been conjectured that the observed large branching fraction of $D_s^+\to\rho^+\eta^\prime$ at the value of $(12.2\pm2.0)\%$ by the CLEO experiment~\cite{CLEO} was overestimated and problematic (see {\it e.g.}, Ref.~\cite{Cheng:2010ry}).  The updated measurement of this mode by BES-III is $(5.80\pm1.46)\%$~\cite{etaetapr}, significantly smaller than the previous one.

\squeezetable
\renewcommand{\arraystretch}{0.85}
\begin{table}[tp!]
\caption{Flavor amplitude decompositions, experimental branching fractions, and predicted branching fractions for the Cabibbo-favored $D \to VP$ decays.  Here $\sf\equiv\sin\!\phi$, $\cf\equiv\cos\!\phi$ and $Y_{sd}\equiv V_{cs}^*V_{ud}$.  The columns of ${\cal B}_{\rm theory}(A1)$ and ${\cal B}_{\rm theory}(S4)$ give our predictions based on Solutions (A1) and (S4) to be shown later in Tables~\ref{globalA} and \ref{globalS}.  For comparison, the columns of ${\cal B}$(pole) and ${\cal B}$(FAT[mix]) are predictions made in Ref.~\cite{Li:2013xsa} based on the pole model and the factorization-assisted topological-amplitude (FAT) approach with the $\rho$-$\omega$ mixing, respectively.  All branching ratios are quoted in units of \%.}
\vspace{6pt}
\begin{ruledtabular}
\begin{tabular}{l l l c c c c c c c c}
Meson & Mode & Representation
& ${\cal B}_{\rm exp}$& ${\cal B}_{\rm theory}(A1)$  & ${\cal B}_{\rm theory}(S4)$  & ${\cal B}$(pole) & ${\cal B}$(FAT[mix]) \\
\hline
$D^0$&$K^{*-}\,\pi^+$          & $Y_{sd}(T_V + E_P)$                                               &$5.43\pm0.44$                    & $5.45\pm0.64$ &$5.43\pm0.70$         &$3.1\pm1.0$         &6.09\\
&$K^-\,\rho^+$                      & $Y_{sd}(T_P + E_V)$                                                &$11.1 \pm 0.9$                   & $11.3\pm2.70$ &$11.4\pm2.78$          &$8.8\pm2.2$         &9.6\\
&$\ol{K}^{*0}\,\pi^0$             &$\frac{1}{\sqrt{2}}Y_{sd}(C_P - E_P)$                         &$3.75\pm0.29$                   & $3.72\pm0.49$ &$3.72\pm0.50$         &$2.9\pm1.0$         &3.25\\
&$\ol{K}^0\,\rho^0$              & $\frac{1}{\sqrt{2}}Y_{sd}(C_V - E_V)$                        &$1.28^{+0.14}_{-0.16}$      &$1.30\pm0.78$ &$1.31\pm0.23$          &$1.7\pm0.7$         &1.17\\
&$\ol{K}^{*0}\,\eta$               & $Y_{sd}({1\over \sqrt{2}}(C_P + E_P)\cf - E_V\sf\,)$&$0.96\pm0.30$                    &$0.92\pm0.36$  &$0.82\pm0.34$        &$0.7\pm0.2$         &0.57\\
&$\ol{K}^{*0}\,\eta\,'$            &$-Y_{sd}({1\over \sqrt{2}}(C_P + E_P)\sf + E_V\cf\,)$&$<0.11$                               &$0.003\pm0.002$ &$0.006\pm0.002$ &$0.016\pm0.005$&0.018\\
&$\ol{K}^0\,\omega$             & $-\frac{1}{\sqrt{2}}Y_{sd}(C_V + E_V)$                       &$2.22 \pm 0.12$                &$2.24\pm0.84$ &$2.24\pm0.29$         &$2.5\pm0.7$        &2.22\\
&$\ol{K}^0\,\phi$                   &$-Y_{sd}E_P$                                                               &$0.847^{+0.066}_{-0.034}$&$0.848\pm0.050$&$0.850\pm0.050$ &$0.80\pm0.2$      &0.800\\
\hline
$D^+$ &$\ol{K}^{*0}\,\pi^+$  & $Y_{sd}(T_V + C_P)$                                                 &$1.57\pm0.13$                   &$1.57\pm0.25$ &$1.57\pm0.25$        &$1.4\pm1.3$  &1.70\\
&$\ol{K}^0\,\rho^+$               & $Y_{sd}(T_P + C_V)$                                                 &$12.08^{+1.20}_{-0.68}$    &$12.15\pm11.69$ &$12.03\pm41.92$&$15.1\pm3.8$ &6.0\\
\hline
$D_s^+$&$\ol{K}^{*0}\,K^+$& $Y_{sd}(C_P + A_V)$                                                  &$3.92\pm0.14$                   &$3.92\pm1.13$ &$3.93\pm1.00$       &$4.2\pm1.7$ &4.07\\
&$\ol{K}^0\,K^{*+}$              & $Y_{sd}(C_V + A_P)$                                                  &$5.4 \pm 1.2$                     &$4.38\pm1.19$ &$3.11\pm1.49$       &$1.0\pm0.6$ &3.1\\
&$\rho^+\,\pi^0$                   & $\frac{1}{\sqrt{2}}Y_{sd}(A_P - A_V)$                          &---                                       &$0.021\pm0.087$&$0.022\pm0.082$&$0.4\pm0.4$ &0\\
&$\rho^+\,\eta$                    & $-Y_{sd}({1\over\sqrt{2}}( A_P + A_V)\cf-T_P\sf)$        &$8.9 \pm 0.8$                     &$8.85\pm1.69$&$8.93\pm3.12$       &$8.3\pm1.3$ &8.8\\
&$\rho^+\,\eta\,'$                 & $Y_{sd}({1\over\sqrt{2}}( A_P + A_V)\sf+T_P\cf)$        &$5.80\pm1.46\footnotemark[1]$&$2.75\pm0.46$ &$2.89\pm0.86$&$3.0\pm0.5$ &1.6\\
 &$\pi^+\,\rho^0$                 & $\frac{1}{\sqrt{2}}Y_{sd}(A_V - A_P)$                           &$0.020\pm0.012$              &$0.021\pm0.087$ &$0.022\pm0.082$&$0.4\pm0.4$ &0.004\\
&$\pi^+\,\omega$                &$\frac{1}{\sqrt{2}}Y_{sd}(A_V + A_P)$                           &$0.24\pm 0.06$                 &$0.24\pm0.15$ &$0.24\pm0.14$        &0                   &0.26\\
&$\pi^+\,\phi$                      &$Y_{sd}T_V$                                                                 &$4.5 \pm 0.4$                     &$4.49\pm0.40$ &$4.51\pm0.43$        &$4.3\pm0.6$ &3.4\\
\end{tabular}
\label{VPCF}
\footnotetext[1]{Data from Ref.~\cite{etaetapr}.}
\end{ruledtabular}
\end{table}
\renewcommand{\arraystretch}{0.85}

\squeezetable
\begin{table}[tp!]
\caption{Same as Table~\ref{VPCF} except for the singly Cabibbo-suppressed decays, $Y_d\equiv V_{cd}^*V_{ud}$ and $Y_s\equiv V_{cs}^*V_{us}$.  All branching ratios are quoted in units of $10^{-3}$.}
\begin{ruledtabular}
\begin{tabular}{l l l c c c c c c c }
Meson & Mode & Representation & ${\cal B}_{\rm exp}$
 &${\cal B}_{\rm theory}(A1)$ &${\cal B}_{\rm theory}(S4)$ & ${\cal B}$(pole) & ${\cal B}$(FAT[mix]) \\
\hline
$D^0$
& $\pi^+\,\rho^-$       & $Y_d(T_V'+E_P')$                                         & $5.09\pm 0.34$&$3.61\pm0.43$ &$4.76\pm0.61$ &$3.5\pm0.6$  &4.66\\
& $\pi^-\,\rho^+$       & $Y_d(T_P'+E_V')$                                         &$10.0\pm 0.6$    &$8.73\pm2.09$ &$8.82\pm2.15$&$10.2\pm1.5$&10.0\\
& $\pi^0\,\rho^0$      & $\frac12Y_d(C_P'+C_V'-E_P'-E_V')$            &$3.82\pm 0.29$  &$3.06\pm0.63$ &$3.90\pm1.62$ &$1.4\pm0.6$ &3.83\\
& $K^+\,K^{*-}$        & $Y_s(T'_V+E'_P)$                                          &$1.62\pm0.15$  &$1.84\pm0.22$ &$1.83\pm0.24$ &$1.6\pm0.3$ &1.73\\
& $K^-\,K^{*+}$        & $Y_s(T'_P+E'_V)$                                          &$4.50\pm0.30$  &$4.44\pm1.07$ &$3.39\pm0.83$ &$4.7\pm0.8$ &4.37\\
& $K^0\,\ol{K}^{*0}$ & $Y_sE'_P+Y_dE'_V$                                     &$<1.5$              &$1.374\pm0.361$ &$1.028\pm0.430$ &$0.16\pm0.05$ &1.1\\
& $\ol{K}^0\,K^{*0}$ & $Y_sE'_V+Y_dE'_P$                                     & $<0.54$           &$1.374\pm0.361$ &$1.028\pm0.430$ &$0.16\pm0.05$ &1.1\\
& $\pi^0\,\omega$   & $\frac{1}{2}Y_d(C'_V-C'_P+E'_P+E'_V)$       & $0.117\pm0.035$ \footnotemark[1]&$0.043\pm0.156$ &$0.272\pm1.509$ &$0.08\pm0.02$ &0.18\\
& $\pi^0\,\phi$         & $\frac{1}{\sqrt{2}}Y_sC'_P$                            &$1.35\pm0.10$ &$0.77\pm0.14$ &$0.66\pm0.11$ &$1.0\pm0.3$ &1.11\\
& $\eta\,\omega$    & $Y_d{1\over 2}(C'_V+C'_P+E'_V+E'_P)\cf$   &$2.21\pm0.23$ \footnotemark[2]&$2.09\pm0.49$ &$2.67\pm2.54$ &$1.2\pm0.3$ &2.0\\
&                            & $-Y_s{1\over\sqrt{2}}C'_V\sf$\\
& $\eta\,' \omega$& $-Y_d{1\over 2}(C'_V+C'_P+E'_V+E'_P)\sf$    &---                      &$0.012\pm0.012$ &$0.046\pm0.067$&$0.0001\pm0.0001$ &0.02\\
&                            & $-Y_s{1\over\sqrt{2}}C'_V\cf$ \\
& $\eta\,\phi$         & $Y_s({1\over\sqrt{2}}C'_P\cf-(E'_V+E'_P)\sf)$&$0.14\pm0.05$ &$0.29\pm0.12$ &$0.29\pm0.08$         &$0.23\pm0.06$ &0.18\\
& $\eta\,\rho^0$     & $-Y_d{1\over 2}(C'_V-C'_P-E'_V-E'_P)\cf$      &   ---                   &$0.60\pm0.40$ &$0.80\pm2.63$        &$0.05\pm0.01$ &0.45\\
&                            & $+Y_s{1\over\sqrt{2}}C'_V\sf$ \\
& $\eta\,' \rho^0$  & $Y_d{1\over 2}(C'_V-C'_P-E'_V-E'_P)\sf$        & ---                     &$0.055\pm0.021$ &$0.105\pm0.075$&$0.08\pm0.02$ &0.27\\
&                           & $+Y_s{1\over\sqrt{2}}C'_V\cf$ \\
\hline
$D^+$
& $\pi^+\,\rho^0$   & $\frac{1}{\sqrt{2}}Y_d(T'_V+C'_P-A'_P+A'_V)$                                        & $0.84\pm 0.15$&$0.51\pm0.28$ &$0.68\pm0.35$ &$0.8\pm0.7$ &0.58\\
& $\pi^0\,\rho^+$   & $\frac{1}{\sqrt{2}}Y_d(T'_P+C'_V+A'_P-A'_V)$                                        &---                      &$4.35\pm5.01$ &$4.27\pm16.51$&$3.5\pm1.6$ &2.5\\
& $\pi^+\,\omega$& $\frac{1}{\sqrt{2}}Y_d(T'_V+C'_P+A'_P+A'_V)$                                        &$0.279\pm0.059$ \footnotemark[1] &$0.165\pm0.269$    &$0.208\pm0.240$ &$0.3\pm0.3$ &0.80\\
& $\pi^+\,\phi$      & $Y_sC'_P$                                                                                                &$5.66^{+0.19}_{-0.21}$&$3.92\pm0.69$ &$3.37\pm0.59$&$5.1\pm1.4$ &5.65\\
& $\eta\,\rho^+$   & $-Y_d{1\over\sqrt{2}}(T'_P+C'_V+A'_V+A'_P)\cf$                &  $<6.8 \footnotemark[3]$                     &$1.43\pm4.60$ &$0.95\pm10.05$&$0.4\pm0.4$ &2.2\\
&                         & $+Y_sC'_V\sf$                \\
& $\eta\,' \rho^+$& $Y_d{1\over\sqrt{2}}(T'_P+C'_V+A'_V+A'_P)\sf$                   & $<5.2 \footnotemark[3]$                     &$0.964\pm0.168$ &$0.958\pm0.507$ &$0.8\pm0.1$ &0.8\\
&                        & $+Y_sC'_V\cf$           \\
& $K^+\,\ol{K}^{*0}$ & $Y_dA'_V+Y_sT'_V$                                                                            &$3.84^{+0.14}_{-0.23}$&$4.00\pm0.82$  &$3.86\pm0.78$   &$4.1\pm1.0$ &3.60\\
& $\ol{K}^0\,K^{*+}$ & $Y_dA'_P+Y_sT'_P$                                                                            &$34\pm16$                   &$14.45\pm2.45$ &$10.03\pm2.62$&$12.4\pm2.4$ &11\\
\hline
$D_s^+$
& $\pi^+\,K^{*0}$& $Y_dT'_V+Y_sA'_V$                                                                                  &$2.13\pm0.36$    &$3.51\pm0.72$ &$3.76\pm0.76$ &$1.5\pm0.7$ &2.35\\
& $\pi^0\,K^{*+}$& $\frac{1}{\sqrt{2}}(Y_dC'_V-Y_sA'_V)$                                                       & ---                        &$1.47\pm0.45$ &$1.04\pm0.48$ &$0.1\pm0.1$ &1.0\\
& $K^+\,\rho^0$ & $\frac{1}{\sqrt{2}}(Y_dC'_P-Y_sA'_P)$                                                        &$2.5\pm0.4$       &$1.58\pm0.38$  &$2.07\pm0.57$ &$1.0\pm0.6$ &2.5\\
& $K^0\,\rho^+$ & $Y_dT'_P+Y_sA'_P$                                                                                   &---                       &$11.25\pm1.90$ &$11.45\pm2.99$&$7.5\pm2.1$ &9.6\\
& $\eta\,K^{*+}$ & $-{1\over\sqrt{2}}(Y_dC'_V+Y_sA'_V)\cf$            &---                        &$0.59\pm2.26$ &$0.64\pm6.09$&$1.0\pm0.4$ &0.2\\
&                        & $+Y_s(T'_P+C'_V+A'_P)\sf$  \\
& $\eta\,' K^{*+}$& ${1\over\sqrt{2}}(Y_dC'_V+Y_sA'_V)\sf$          &  ---                      &$0.42\pm0.15$ &$0.32\pm0.14$ &$0.6\pm0.2$ &0.2\\
&                         & $+Y_s(T'_P+C'_V+A'_P)\cf$\\
& $K^+\,\omega$ & $\frac{1}{\sqrt{2}}\left(Y_dC'_P+Y_sA'_P\right)$                                      &$<2.4$                  &$1.05\pm0.34$ &$2.15\pm0.56$ &$1.8\pm0.7$ &0.07\\
& $K^+\,\phi$       & $Y_s(T'_V+C'_P+A'_V)$                                                                          & $0.164\pm0.041$&$0.111\pm0.060$  &$0.112\pm0.068$ &$0.3\pm0.3$ &0.166\\
\end{tabular}
\label{VPSCS}
\footnotetext[1]{Data from Ref.~\cite{BESIII}.}
\footnotetext[2]{Data from Ref.~\cite{kass}.}
\footnotetext[3]{Data from Ref.~\cite{CLEO}.}
\end{ruledtabular}
\end{table}

\renewcommand{\arraystretch}{0.85}
\squeezetable
\begin{table}
\caption{Same as Table~\ref{VPCF} except for the doubly Cabibbo-suppressed decays and $Y_{ds}\equiv V_{cd}^*V_{us}$.  All branching ratios are quoted in units of $10^{-4}$.}
\begin{ruledtabular}
\begin{tabular}{l l l c c c c c c c}
Meson & Mode & Representation & ${\cal B}_{\rm exp}$
&${\cal B}_{\rm theory}(A1)$ &${\cal B}_{\rm theory}(S4)$  & ${\cal B}$(pole) & ${\cal B}$(FAT[mix]) \\
\hline
$D^0$
& $K^{*+}\,\pi^-$            & $Y_{ds}(T''_P+E''_V)$                                            & $3.45^{+1.80}_{-1.02}$&$3.77\pm0.90$ &$2.88\pm0.70$&$2.7\pm0.6$ &4.72\\
& $K^{*0}\,\pi^0$           & $\frac{1}{\sqrt{2}}Y_{ds}\left( C''_P-E''_V \right)$     &---&  $0.49\pm0.23$ &$0.47\pm0.12$ &$0.8\pm0.3$ &0.9\\
& $\phi\,K^0$                 & $-Y_{ds}E''_V$                                                          & ---&$0.04\pm0.03$ &$0.01\pm0.01$ &$0.20\pm0.06$ &0.2\\
& $\rho^-\,K^+$              & $Y_{ds}(T''_V+E''_P)$                                              &--- &$1.34\pm0.16$ &$1.76\pm0.23$ &$0.9\pm0.3$ &1.5\\
& $\rho^0\,K^0$              & $\frac{1}{\sqrt{2}}Y_{ds}(C''_V-E''_P)$                     &--- &$1.06\pm0.38$ &$1.30\pm1.80$ &$0.5\pm0.2$ &0.3\\
& $\omega\,K^0$            & $-\frac{1}{\sqrt{2}}Y_{ds}(C''_V+E''_P)$                    &--- &$0.40\pm0.37$ &$0.61\pm1.74$&$0.7\pm0.2$ &0.6\\
& $K^{*0}\,\eta$               & $Y_{ds}({1\over\sqrt{2}}(C''_P+E''_V)\cf-E''_P\sf)$ &--- &$0.53\pm0.10$ &$0.46\pm0.08$ &0.08 &0.2\\
& $K^{*0}\,\eta^{\prime}$ & $Y_{ds}({1\over\sqrt{2}}(C''_P+E''_V)\sf+E''_P\cf)$& ---& $0.001\pm0.0004$ &$0.002\pm0.001$ &$0.004\pm0.001$ &0.005\\
\hline
$D^+$
& $K^{*0}\,\pi^+$             & $Y_{ds}(C''_P+A''_V)$                                             &$3.9\pm0.6$  & $2.94\pm0.85$ &$2.66\pm0.68$ &$2.2\pm0.9$ &3.33\\
& $K^{*+}\,\pi^0$             & $\frac{1}{\sqrt{2}}Y_{ds}\left( T''_P-A''_V \right)$      &  ---& $5.76\pm0.85$ &$3.98\pm1.17$ &$4.0\pm0.9$ &3.9\\
& $\phi\,K^+$                   & $Y_{ds}A''_V$                                                          & --- & $0.02\pm0.02$ &$0.02\pm0.01$ &$0.2\pm0.2$ &0.02\\
& $\rho^+\,K^0$               & $Y_{ds}(C''_V+A''_P)$                                             & --- & $2.81\pm0.76$ &$2.39\pm1.14$ &$0.5\pm0.4$ &3.3\\
& $\rho^0\,K^+$               & $\frac{1}{\sqrt{2}}Y_{ds}(T''_V-A''_P)$                     & $2.1\pm0.5$&$1.66\pm0.24$&$2.09\pm0.44$ &$0.5\pm0.4$ &2.4\\
& $\omega\,K^+$             & $\frac{1}{\sqrt{2}}Y_{ds}(T''_V+A''_P)$                    &--- & $0.95\pm0.20$ &$1.90\pm0.42$ &$1.8\pm0.5$ &0.7\\
& $K^{*+}\,\eta$               & $-Y_{ds}({1\over\sqrt{2}}(T''_P+A''_V)\cf-A''_P\sf)$   &  ---& $1.89\pm0.40$ &$1.33\pm0.33$ &$1.4\pm0.2$ &1.0\\
& $K^{*+}\,\eta^{\prime}$ & $Y_{ds}({1\over\sqrt{2}}(T''_P+A''_V)\sf+A''_P\cf)$  &  ---& $0.02\pm0.01$ &$0.02\pm0.01$ &$0.020\pm0.007$ &0.01\\
\hline
$D^+_s$
& $K^{*+}\,K^0$ & $Y_{ds}(T''_P+C''_V)$                                                             &--- &$1.55\pm1.49$ &$1.29\pm4.48$ &$2.3\pm0.6$ &1.1\\
& $K^{*0}\,K^+$ & $Y_{ds}(T''_V+C''_P)$                                                             & $0.90\pm0.51$&$0.17\pm0.03$ &$0.19\pm0.03$ &$0.2\pm0.2$ &0.23\\
\end{tabular}
\label{VPDCS}
\end{ruledtabular}
\end{table}

\section{Formalism}\label{sec:formalism}

Our convention of the quark contents for light pseudoscalar mesons are $\pi^+=u\ol d$, $\pi^0=(d\ol d-u\ol u)/\sqrt2$, $\pi^-=-d\ol u$, $K^+=u\ol s$, $K^0=d\ol s$, $\ol K^0=s\ol d$, $K^-=-s\ol u$ while those for light vector mesons are $\rho^+=u\ol d$, $\rho^0=(d\ol d-u\ol u)/\sqrt2$, $\rho^-=-d\ol u$, $K^{*+}=u\ol s$, $K^{*0}=d\ol s$, $\ol K^{*0}=s\ol d$, $K^{*-}=-s\ol u$, $\omega=(u\ol u+d\ol d)/\sqrt2$ and $\phi=s\ol s$.  The physical states of $\eta$ and $\eta^\prime$ in terms of the quark-flavor ones $\eta_q =\frac{1}{\sqrt2}(u\ol u+d\ol d)$ and $\eta_s =s\ol s$ are given by
\begin{equation}
\begin{pmatrix}
\eta \\
\eta'
\end{pmatrix}
=
\begin{pmatrix}
\cos\phi  & -\sin\phi \\
\sin\phi  &  \cos\phi \\
\end{pmatrix}
\begin{pmatrix}
\eta_q \\
\eta_s
\end{pmatrix} ~,
\end{equation}
with the mixing angle $\phi$ ranging from $39^\circ$ to $49^\circ$.  We shall use the recent LHCb measurement~\cite{etaangle} to fix $\phi$ at $43.5^\circ$ in our numerical calculations.

The partial decay width of the $D$ meson into a vector and a pseudoscalar meson can be expressed in two different ways:
\begin{equation}
\Gamma(D\to VP)=\frac{p_c^3}{8\pi m_D^2}|{\tilde{\cal  M}}|^2 ~,
\label{decaywidthA}
\end{equation}
and
\begin{equation}
\Gamma(D\to VP)=\frac{p_c}{8\pi m_D^2}\sum_{\rm pol.}|{\cal M}|^2 ~,
\label{decaywidthS}
\end{equation}
where $m_D$ is the $D$ meson mass, and $p_c$ is the center-of-mass momentum of either meson in the final state.  Note that the partial widths and thus the branching fractions throughout this paper are $CP$-averaged.  The summation in Eq.~(\ref{decaywidthS}) is over the polarizations of the vector meson.
The branching fraction for a specific decay process can be obtained by multiplying the partial width with the $D$ meson lifetime.  The relation between the amplitudes $\tilde{\cal  M}$ and $\cal M$ is $\tilde{\cal  M}(\epsilon\cdot p_D)=(m_D/m_V){\cal  M}$, where $\epsilon^\mu$ and $m_V$ denote respectively the polarization vector and mass of $V$ meson, and $p_D^\mu$ is the momentum of $D$ meson.

The flavor amplitude decompositions for all the $D\to VP$ decay modes are shown in Tables~\ref{VPCF}--\ref{VPDCS}, in which we have defined the CKM factors $Y_{sd}\equiv V_{cs}^*V_{ud} \sim {\cal O}(1)$, $Y_d\equiv V_{cd}^*V_{ud} \sim {\cal O}(\lambda)$, $Y_s\equiv V_{cs}^*V_{us} \sim {\cal O}(\lambda)$, and $Y_{ds}\equiv V_{cd}^*V_{us} \sim {\cal O}(\lambda^2)$ for simplicity.  To a very good approximation, the involved four CKM matrix factors only depend on the Wolfenstein parameter $\lambda$, which is fixed to $0.22543$~\cite{CKMfitter} by neglecting its small uncertainty.

With the SU(3)$_F$ symmetry in the diagrammatic approach, we only need four types of amplitudes for all the $D\to VP$ decays: the color-allowed amplitude $T$, the color-suppressed amplitude $C$, the $W$-exchange amplitude $E$, and the $W$-annihilation amplitude $A$.
We associate a subscript $P$ or $V$ to each flavor amplitude, {\it e.g.}, $T_{P,V}$, to denote the amplitude in which the spectator quark goes to the pseudoscalar or vector meson in the final state.  These two kinds of amplitudes do not have any obvious relation {\it a priori}.

Here we briefly comment on the branching fraction of the $D_s^+\to\rho^+\eta'$ mode recently reported by the BES-III Collaboration~\cite{etaetapr}.  Its central value, seen to deviate from theory predictions, can be constrained using two related modes.  From the flavor decompositions in Table~\ref{VPCF}, one derives a sum rule:
\begin{equation}
\frac{1}{\sf}{\cal A}(D_s^+\to\pi^+\omega)
= \frac{\cf}{\sf}{\cal A}(D_s^+\to\rho^+\eta)+{\cal A}(D_s^+\to\rho^+\eta') ~,
\label{sumrule}
\end{equation}
where $s_\phi\equiv\sin\phi$ and $c_\phi\equiv\cos\phi$.
Taking the current data of ${\cal B}(D_s^+\to\pi^+\omega)$ and ${\cal B}(D_s^+\to\rho^+\eta)$ and noting a simple triangular inequality, we obtain the bounds $(2.19\pm0.27)\%<{\cal B}(D_s^+\to\rho^+\eta')<(4.51\pm0.38)\%$, consistent with the current data within the $1\sigma$ level.

The decay $D_s^+\to\rho^0\pi^+$ plays a crucial role in determining the annihilation amplitudes $A_{P,V}$ in the current analysis.  It is so because this is the only observed mode whose $A_P$ and $A_V$ have opposite signs, while others involve their sum.  Without this observable, both the magnitudes and the strong phases of $A_{P,V}$ cannot be settled.  Before 2010, this mode was quoted by the PDG as ``not seen.''  A Dalitz-plot analysis of $D_s^+\to\pi^+\pi^+\pi^-$ by BaBar yielded the fit fraction $\Gamma(D_s^+\to \rho^0\pi^+)/\Gamma(D_s^+\to\pi^+\pi^+\pi^-)= (1.8\pm0.5\pm1.0)\%$~\cite{BaBarrhopi}.  Given the branching fraction ${\cal B}(D_s^+\to\pi^+\pi^+\pi^-)=(1.09\pm0.05)\%$ \cite{pdg}, the BaBar result leads to ${\cal B}(D_s^+\to \rho^0\pi^+)=(2.0\pm1.2)\times 10^{-4}$.

\section{Data Fitting}\label{sec:fitting}

\begin{table}[tp!]
\caption{Fit results using Eq.~(\ref{decaywidthA}) and $\phi=43.5^\circ$.  The amplitude sizes are quoted in units of $10^{-6}$, and the strong phases in units of degrees.  Only those solutions which can sufficiently well accommodate the singly Cabibbo-suppressed modes are shown.
}
\begin{ruledtabular}
\begin{tabular}{l l l c c c c c c c c}
         &$|T_V|$                   &$|T_P|$            &$\delta_{T_P}$          &$|C_V|$                          &$\delta_{C_V}$          &$|C_P|$                  &$\delta_{C_P}$                &$|E_V|$                         &$\delta_{E_V}$     \\
          &$|E_P|$                            &$\delta_{E_P}$   &$|A_P|$                           &$\delta_{A_P}$             &$|A_V|$ &$\delta_{A_V}$ &$\chi^2_{min}$ &quality\\
\hline
(A1)  &$4.21^{+0.18}_{-0.19}$&$8.46^{+0.22}_{-0.25}$&$57^{+35}_{-41}$&$4.09^{+0.16}_{-0.25}$&$-145^{+29}_{-39}$&$4.08^{+0.37}_{-0.36}$&$-157\pm2$    &$1.19^{+0.64}_{-0.46}$&$-85^{+42}_{-39}$  \\
         &$3.06\pm0.09$&$98\pm5$     &$0.64^{+0.14}_{-0.27}$&$152^{+48}_{-50}$          &$0.52^{+0.24}_{-0.19}$  &$122^{+70}_{-42}$          &$5.22$ &$0.0223$\\
           \hline
(A2)  &$4.26^{+0.18}_{-0.19}$&$8.13^{+0.61}_{-0.47}$&$69^{+30}_{-56}$&$4.20\pm0.12$&$-82^{+36}_{-26}$ &$4.34^{+0.41}_{-0.40}$&$-158\pm2$    &$0.61^{+0.78}_{-0.12}$&$-90^{+78}_{-60}$ \\
            &$3.06\pm0.09$&$100\pm5$  &$0.71^{+0.08}_{-0.36}$&$-32^{+64}_{-82}$          &$0.40^{+0.35}_{-0.10}$  &$-42^{+99}_{-55}$          &$6.23$ &$0.0126$\\
            \hline
(A3)  &$4.26^{+0.17}_{-0.18}$&$8.43^{+0.24}_{-0.53}$&$34^{+87}_{-40}$&$4.07^{+0.22}_{-0.42}$&$-168^{+154}_{-26}$&$4.36^{+0.32}_{-0.34}$&$-158\pm2$     &$1.26^{+0.92}_{-0.72}$&$-106^{+43}_{-37}$\\
           &$3.06\pm0.09$&$100\pm5$    &$0.53^{+0.25}_{-0.21}$&$-79^{+64}_{-32}$          &$0.62^{+0.16}_{-0.30}$  &$-48^{+60}_{-31}$          &$7.25$ &$0.0071$\\
            \hline
(A4)  &$4.21^{+0.18}_{-0.19}$ &$8.01^{+0.52}_{-0.58}$&$31^{+26}_{-57}$&$4.20^{+0.13}_{-0.16}$&$-119^{+34}_{-107}$&$4.06^{+0.44}_{-0.50}$&$-157\pm2$   &$0.66^{+0.51}_{-0.17}$&$-96\pm79$ \\
          &$3.06\pm0.09$&$98^{+5}_{-6}$     &$0.61^{+0.16}_{-0.25}$&$156^{+55}_{-50}$          &$0.54^{+0.21}_{-0.22}$  &$123^{+125}_{-48}$          &$7.98$ &$0.0047$\\
           \hline
(A5)  &$3.84\pm0.17$&$8.48^{+0.21}_{-0.25}$&$-54^{+28}_{-23}$&$4.09^{+0.17}_{-0.27}$&$104^{+28}_{-23}$&$5.00^{+0.10}_{-0.12}$&$-165^{+2}_{-3}$     &$1.22^{+0.66}_{-0.47}$&$164^{+25}_{-27}$ \\
           &$3.03\pm0.09$&$-85\pm4$   &$0.43^{+0.13}_{-0.09}$&$30^{+29}_{-34}$    &$0.76^{+0.07}_{-0.10}$  &$18\pm19$  &$14.24$ &$0.0002$\\  
\end{tabular}
\label{globalA}
\end{ruledtabular}
\label{fitting}
\end{table}

\begin{table}[tp!]
\caption{Same as Table~\ref{fitting} except that Eq.~(\ref{decaywidthS}) is employed for the fit. The amplitude sizes are quoted in units of $10^{-6}(\epsilon\cdot p_D)$.}
\begin{ruledtabular}
\begin{tabular}{l l l c c c c c c c c}
         &$|T_V|$                     &$|T_P|$        &$\delta_{T_P}$          &$|C_V|$                          &$\delta_{C_V}$   &$|C_P|$                  &$\delta_{C_P}$                &$|E_V|$                            &$\delta_{E_V}$        \\
         &$|E_P|$                         &$\delta_{E_P}$          &$|A_P|$                           &$\delta_{A_P}$             &$|A_V|$ &$\delta_{A_V}$ &$\chi^2_{min}$ &quality\\
\hline
(S1)  &$2.19\pm0.09$&$3.40^{+0.17}_{-0.18}$&$57^{+30}_{-53}$&$1.76^{+0.05}_{-0.09}$&$-94^{+36}_{-28}$&$2.09^{+0.11}_{-0.17}$&$-159\pm1$     &$0.27^{+0.34}_{-0.07}$&$-116^{+77}_{-58}$  \\
         &$1.67\pm0.05$&$108\pm4$    &$0.26^{+0.06}_{-0.11}$&$-31^{+65}_{-59}$          &$0.20^{+0.10}_{-0.07}$  &$-1^{+68}_{-58}$          &$5.558$ &$0.0184$\\ 
            \hline
(S2)  &$2.19\pm0.09$ &$3.40^{+0.16}_{-0.19}$&$64^{+30}_{-60}$&$1.76^{+0.05}_{-0.09}$&$-88^{+35}_{-26}$ &$2.10^{+0.11}_{-0.17}$&$-159\pm1$  &$0.28^{+0.33}_{-0.07}$&$-114^{+78}_{-61}$  \\
          &$1.67\pm0.05$&$108\pm4$    &$0.26^{+0.05}_{-0.12}$&$-23^{+63}_{-68}$          &$0.20^{+0.10}_{-0.07}$  &$6^{+71}_{-66}$          &$5.564$ &$0.0183$\\ 
            \hline
(S3) &$2.17^{+0.09}_{-0.10}$&$3.47^{+0.11}_{-0.34}$&$33^{+47}_{-28}$&$1.75^{+0.06}_{-0.10}$&$-172^{+26}_{-37}$ &$2.03^{+0.18}_{-0.17}$&$-159\pm1$  &$0.39^{+0.29}_{-0.17}$&$-123^{+46}_{-117}$  \\
           &$1.67\pm0.05$&$107^{+5}_{-4}$    &$0.23^{+0.07}_{-0.09}$&$109^{+46}_{-51}$          &$0.23^{+0.07}_{-0.09}$  &$77^{+47}_{-50}$          &$5.90$ &$0.0152$\\ 
            \hline
(S4)  &$2.18^{+0.11}_{-0.10}$ &$3.38^{+0.27}_{-0.28}$&$9^{+83}_{-82}$&$1.77\pm0.05$&$-142^{+81}_{-147}$ &$2.06^{+0.17}_{-0.19}$&$-159^{+1}_{-2}$  &$0.25^{+0.18}_{-0.05}$&$-146^{+65}_{-114}$  \\
           &$1.67\pm0.05$&$108\pm5$   &$0.19^{+0.10}_{-0.07}$&$100^{+51}_{-79}$          &$0.26^{+0.05}_{-0.10}$  &$72^{+45}_{-38}$          &$8.08$ &$0.0045$\\ 
            \hline
(S5)  &$1.81\pm0.11$ &$3.50^{+0.10}_{-0.11}$&$-32^{+34}_{-25}$&$1.73^{+0.06}_{-0.09}$&$125^{+35}_{-26}$ &$2.25^{+0.04}_{-0.05}$&$-162^{+2}_{-3}$     &$0.46^{+0.24}_{-0.17}$&$-179^{+35}_{-33}$\\
            &$1.65\pm0.05$&$-86\pm4$ &$0.17^{+0.05}_{-0.03}$&$30^{+28}_{-31}$          &$0.31^{+0.03}_{-0.04}$  &$20^{+18}_{-17}$          &$33.78$ &$0.0000$\\ 
             \hline
(S6)  &$1.81^{+0.12}_{-0.11}$ &$3.50^{+0.10}_{-0.11}$&$-34^{+37}_{-23}$&$1.73^{+0.06}_{-0.09}$&$122^{+33}_{-24}$&$2.25^{+0.04}_{-0.05}$&$-162^{+2}_{-3}$    &$0.46^{+0.24}_{-0.17}$&$179^{+37}_{-31}$  \\
            &$1.64\pm0.05$&$-86\pm4$ &$0.17^{+0.05}_{-0.03}$&$29^{+29}_{-31}$          &$0.31^{+0.03}_{-0.04}$  &$19^{+19}_{-16}$          &$33.79$ &$0.0000$\\ 
\end{tabular}
\label{globalS}
\end{ruledtabular}
\end{table}

Since the measured CP asymmetries are consistent with zero for most of the $D\to VP$ channels, we only take into account the branching fractions in our fit.  We start exclusively with the Cabibbo-favored decay modes, and will test the flavor SU(3) symmetry by using the fit results to predict the branching fractions of Cabibbo-suppressed decays.  There are totally 16 observables with 15 theory parameters as shown in Table~\ref{VPCF}.  We assume no correlations among the theory parameters.  By performing a $\chi^2$ fit to data, we extract the magnitude and strong phase of each flavor diagram.  We have found many possible solutions with local $\chi^2$ minima.  Some of them are not well-separated by sufficiently high ``$\chi^2$ barriers'' to render good 1-$\sigma$ ranges.  In Table~\ref{globalA} and Table~\ref{globalS}, we only present those whose predicted branching fractions for  singly Cabibbo-suppressed modes have better agreement with data.  In particular, in the effort of discarding irrelevant solutions, the $D^0\to\pi^0\omega$ mode plays a major role.  To obtain the 1-$\sigma$ range of each theory parameter, we enable the other parameters to vary freely around their best-fit values and minimize the $\chi^2$ value until the change in $\chi^2$, $\Delta\chi^2$, reaches $1$.  In some rare cases when the $\chi^2$ barrier is not sufficiently high to separate two local minima, we stop the 1-$\sigma$ range scan at the obvious boundary.

Solutions (A) and (S) are obtained when the invariant decay amplitude of $D\to VP$ is extracted using  Eq.~(\ref{decaywidthA}) and Eq.~(\ref{decaywidthS}), respectively.  Note that although the amplitudes derived from them are related to each other, corresponding solutions in set (A) and set (S) have similar but not exactly the same strong phases, as they contain different factors of final-state meson mass [as seen from the relation $\tilde{\cal  M}(\epsilon\cdot p_D)=(m_D/m_V){\cal  M}$].  Since what are fitted are the branching fractions, there are degeneracies in the $\chi^2$ value when all the strong phases simultaneously flip signs or change by $180^\circ$.  We list only one of them in the tables.

In general, the uncertainties associated with certain strong phases are relatively large in some of the solutions.  Usually, the size of the associated amplitude uncertainty is also bigger.  Among all the theory parameters, the uncertainties associated with $|E_P|$, $\delta_{E_P}$ and $\delta_{C_P}$ are much smaller than the others.  In addition, their best-fit values are quite stable across different solutions.  The $D^0\to \ol K^0\phi$ and $D_s^+\to\pi^+\phi$ decays are solely governed by $E_P$ and $T_V$, respectively.  They hence play a dominant role in fixing the sizes of these two flavor amplitudes and their associated errors.  As alluded to earlier, the recently measured branching fraction of $D_s^+\to\pi^+\rho^0$ helps fixing the magnitudes and strong phases of the annihilation amplitudes $A_{P,V}$ for the first time, although their uncertainties, especially in the strong phases, are still large.

The flavor amplitudes generally respect the following hierarchy pattern: $|T_P|>|T_V|\sim|C_{P,V}|>|E_P|>|E_V|\sim|A_{P,V}|$.  Because of the different momentum $p_c$ dependence in Eqs.~(\ref{decaywidthA}) and (\ref{decaywidthS}), the amplitude sizes in Solution (A) are larger than the counterparts in Solution (S).  Both Solutions (A) and (S) can be divided into two different groups.  The first one includes Solutions (A1)-(A4) [or Solutions (S1)-(S4)] with $\delta_{C_P} \simeq -158^\circ$ and positive $\delta_{E_P}$, while the second group includes Solutions (A5) [or (S5)-(S6)] with $\delta_{C_P} \simeq -165^\circ$ and negative $\delta_{E_P}$.  As a correlation, the values of $|T_P|$, $|C_P|$ and $|A_V|$ increase when going from the first group to the second one, while those of $|T_V|$, $|E_P|$ and $|A_P|$ decrease.  From Table~\ref{VPCF}, it is seen that $|T_P|$ has to be large in order to account for the measured large rates of $D^+\to K^-\rho^+$ and $D^+\to\ol K^0\rho^+$.  The relation $E_V \approx -E_P$ advocated in Ref.~\cite{EPEV} is disfavored by the data.  Rather, we observe that $|E_V|$ is significantly smaller than $|E_P|$.  Though the uncertainties are still large, $A_P$ and $A_V$ are generally one order of magnitude smaller than the tree and color-suppressed amplitudes.
Moreover, in some solutions, the $A$'s are comparable to $E_V$ in magnitude.  Therefore, the contributions of the $W$-annihilated amplitudes $A_{P,V}$ are not negligible.

For both Solutions (A) and (S), the major $\chi^2$ contribution comes from the $D^+_s\to\rho^+\eta'$ mode as the predicted branching fractions for this mode are significantly smaller than the current data.  For Solutions (A5), (S5) and (S6), the predicted ${\cal B}(D_s^+\to\pi^+\phi)$ shows a large deviation from the data, resulting in larger $\chi^2$ values.  Hence, the $D_s^+\to\pi^+\phi$ decay helps distinguishing solutions in the first group [{\it i.e.}, (A1)-(A4) and (S1)-(S4)] from those in the second group.

Measurements of singly Cabibbo-suppressed decay modes are useful in distinguishing different solutions.  In Solution (A1), the predicted ${\cal B}(D^0\to\pi^0\phi)$, ${\cal B}(D^0\to\pi^+\rho^-)$ and ${\cal B}(D_s^+\to\pi^+K^{*0})$ deviate from the data more significantly than the other modes.  Solutions (A2) and (A3) are strongly disfavored by the measurements of $D^0\to\pi^0\omega$ and $D^+\to\pi^+\omega$ as the predicted branching fractions are considerably larger.  Among all these decay modes, the predicted ${\cal B}(D^+\to\pi^+\phi)$  has the largest deviation from the data in Solution (A4).  On the other hand, Solution (A5) is disfavored by the measurements of $D^0\to\pi^0\omega$ and $D^+\to K^+\ol K^{*0}$.  In general, Solution (A1) can explain the current data much better than all the other solutions in (A).

The predicted branching fraction of $D^+\to\pi^+\omega$ ($D^+\to\pi^+\phi$) in Solutions (S1) and (S2) is much larger (smaller) than the measurement.  Hence, these two solutions are disfavored by the current data.  The measurements of ${\cal B}(D^0\to\pi^0\omega)$ and ${\cal B}(D^+\to\pi^+\phi)$ can be used to rule out Solution (S3).  As for Solution (S4), the predicted ${\cal B}(D^+\to\pi^+\phi)$ deviates from the data the most.  For Solutions (S5) and (S6), the predicted ${\cal B}(D^+\to K^+\ol K^{*0})$ has the largest deviation from the data among all the decay modes.  Overall, though Solution (S4) cannot explain ${\cal B}(D^+\to\pi^+\phi)$  very well, the predicted branching fractions for all the other decay modes are much closer to the current data than the rest of solutions in (S).

We note in passing that a fit to only singly Cabibbo-suppressed decay modes had been tried.  Not only did we obtain many more solutions, we also could not obtain results with small $\chi^2$ values.  This reflects the fact that these data present inconsistency within this framework.  This also explains why we choose to use Solution (S4) rather than (S1) although the latter has a lower $\chi^2$ value and is closer to Solution (A1) as far as the strong phases are concerned.

In contrast to  singly Cabibbo-suppressed decay modes, all the solutions can explain the available data of  doubly Cabibbo-suppressed decay modes sufficiently well, as will be discussed further in the next section.  Thus, currently singly Cabibbo-suppressed decays play an essential role in singling out preferred solutions.

Before closing the section, we make a comparison between Solutions (A1) to (A5) obtained in the current work and Solutions (A) and (A') given in Table~VII of Ref.~\cite{Cheng:2010ry}.  In the earlier analysis~\cite{Cheng:2010ry}, the data preferred Solution (A) over Solution (A'), primarily because the former had a larger $|C_P|$ than that of the latter and hence it fits the singly Cabibbo-suppressed modes $\pi^{+,0} \phi$ better.  In the current analysis, we notice that ${\cal B}(\overline{K}^{*0} \pi^0) = (3.75 \pm 0.29) \%$ is significantly larger than the 2010 data of $(2.82 \pm 0.35) \%$.  This change has the effect of enlarging $|C_P|$ of Solution (A') to have a more constructive interference with $E_P$ and giving the current Solutions (A1) to (A4).  Such an identification can be seen by paying attention to the strong phases of $C_P$ and $E_P$.  This also results in a better fit to the $\pi^{+,0} \phi$ modes, which involve purely the $C_P$ amplitude.  In contrast, the previously favored Solution (A) evolves to the current Solution (A5) with a smaller $|C_P|$ than before.  A comparison between Solutions of type (S) can be made analogously, and one would find the correspondence between Solutions (S1) to (S4) to Solution (S') and Solutions (S5) to (S6) to Solution (S').

It is also noted that $|C_P|$ and $|C_V|$ are comparable in Solutions (A1) to (A4), but have a small hierarchy in Solutions (S1) to (S4).  As a way to tell whether the amplitudes extracted using Eq.~(\ref{decaywidthA}) or Eq.~(\ref{decaywidthS}) show better flavor symmetry, one can resort to the $D_s \to \overline{K}^{*0} K^+$ decay, governed by $C_P$, and the ${\bar K}^0 K^{*+}$ decay, dominated by $C_V$.  Experimental measurements of the ratio of the branching fractions of them will help us determine to tell the preference of them.  The current data slightly favor (A1) over (S4).  Since the former decay has been measured several times with similar results before and the latter was measured in 1989 \cite{Chen:1989tu}, it is obvious that the ${\bar K^0} K^{*+}$ mode should be updated.

\section{Predictions}\label{sec:predictions}

As explained in the previous section, among all the solutions listed in Tables~\ref{globalA} and \ref{globalS}, Solutions (A1) and (S4) are favored by the current data with the former being slightly preferred after considering all the decay modes, including both singly and doubly Cabibbo-suppressed ones.  We therefore make predictions for all the branching fractions based on Solutions (A1) and (S4) by assuming the SU(3)$_F$ symmetry, with the flavor amplitudes for singly and doubly Cabibbo-suppressed decays being exactly the same as those for  Cabibbo-favored decays ({\it i.e.}, the unprimed, primed, and doubly-primed amplitudes of the same topology are all equal).  In particular, information of the sizes and strong phases of $A_{P,V}$ enables us to predict the branching fractions of the decay modes involving these amplitudes within this framework for the first time.  The results are already given in the columns of ${\cal B}_{\rm theory}(A1)$ and ${\cal B}_{\rm theory}(S4)$ in Tables~\ref{VPCF}, \ref{VPSCS} and \ref{VPDCS}.  One purpose is to test the SU(3)$_F$ symmetry.  Predictions made in the pole model and in the factorization-assisted topological-amplitude (FAT) approach with the $\rho$-$\omega$ mixing~\cite{Li:2013xsa} are also shown in the tables for comparison.

Consider the $D_s^+\to\rho^+\eta$ and $\rho^+\eta'$ decays and Solution (A1).  Since $|T_P| \gg |A_V|, |A_P|$, the color-allowed amplitude $T_P$ is the dominant contribution to the flavor amplitude of the decay mode $D_s^+\to\rho^+\eta^{(\prime)}$. From Table~\ref{VPCF}, once the $W$-annihilation amplitudes are neglected, the ratio of the theoretical branching fraction ${\cal B}(D_s^+\to\rho^+\eta)$ to ${\cal B}(D_s^+\to\rho^+\eta^\prime)$ can simply be parametrized in terms of the mixing angle $\phi$ and the center-of-mass momentum of either meson in the final state
\begin{equation}
{{\cal B}(D_s^+\to \rho^+\eta)\over {\cal B}(D_s^+\to \rho^+\eta')}
\approx \left( {\sin\phi\over \cos\phi}\right)^2
\left({ p_c(D_s\to\rho\eta)\over p_c(D_s\to\rho\eta') }\right)^3
~,
\label{eq:rhoetaratio}
\end{equation}
which numerically is about $3.4$.  This is close to the value of $3.2$, as the central value obtained using Solution (A1) (see Table~\ref{VPCF}) when all the $T_P$ and $A_{P,V}$ contributions are considered.  This is due to the fact that the combination $A_P + A_V$ is roughly perpendicular to $T_P$ in Solution (A1), so that the ratios with and without the $W$-annihilations are roughly the same.  While the predicted ${\cal B}(D_s^+\to\rho^+\eta)$ is close to the CLEO measurement of $(8.9\pm0.8)\%$, the calculated branching fraction of $D_s^+\to\rho^+\eta'$ is substantially below the recent BES-III result of $(5.80\pm1.46)\%$. Indeed, all the existing model calculations yield  around $3\%$ \cite{Cheng:2010ry, Bhattacharya:2008ke,Buccella:1994nf, Li:2013xsa, Zou:2013ez}. If ${\cal B}(D_s^+\to\rho^+\eta')$ still remains to be of order $6\%$ in the future experiments,  this may hint at a sizable flavor-singlet contribution unique to the $\eta_0$ production.  This issue should be clarified both experimentally and theoretically.

Measurements of singly and doubly Cabibbo-suppressed modes serve as a testing ground for our working assumption of flavor SU(3) symmetry.  The predicted branching fractions for the singly Cabibbo-suppressed modes are one order of magnitude smaller than those of the Cabibbo-favored modes due to the suppression of the CKM matrix elements.  Many of the singly Cabibbo-suppressed modes ({\it e.g.}, $D^+\to K^+\ol K^{*0}$ and $D^0\to\eta\omega$) can be nicely explained in the framework of flavor SU(3) symmetry.  The decay amplitudes of $D^0\to K^0\ol K^{*0}$ and $\ol K^0K^{*0}$ both contain $E_V$ and $E_P$, but with different CKM matrix elements.  As both $Y_d$ and $Y_s$ are around 0.2, their predicted branching fractions turn out to be virtually the same.  We note that our prediction  is close to the current upper bound at 90\% confidence level (CL) for $K^0\ol K^{*0}$ and exceeds the upper bound for the $\ol K^0K^{*0}$ mode.  Precise determinations of these observables will determine whether our picture is correct.
The flavor amplitudes involved in the modes $\pi^0\phi$ and $\pi^+\phi$ are the same except the former is suppressed by a factor $1/\sqrt2$. Also, the lifetime of $D^0$ is around 2.5 times shorter than $D^+$. Thus, the branching fraction of $\pi^0\phi$ is expected to be about 5 times smaller than $\pi^+\phi$, as verified by the current data.
The $D^+\to\ol K^0K^{*+}$ and $D_s^+\to K^0\rho^+$ rates are expected to be larger since they are dominated by $T_P$ whose fit value is the largest among all flavor amplitudes.  The current central value of ${\cal B}(D^+\to\ol K^0K^{*+})$ is somewhat too large in comparison with theory predictions, although the error bar is still big.  The predicted ${\cal B}(D^0\to\pi^+\rho^-)$, ${\cal B}(D^0\to \pi^0\phi)$ and ${\cal B}(D_s^+\to\pi^+K^{*0})$ in Solution (A1) deviate from the data more significantly, while the predicted ${\cal B}(D\to\pi\phi)$, ${\cal B}(D^0\to\pi^0\omega)$, ${\cal B}(D_s^+\to\pi^+K^{*0})$ and ${\cal B}(D^0\to K^-K^{*+})$ have larger deviations in Solution (S4).  For ${\cal B}(D_s^+\to\pi^+K^{*0})$, there is a constructive interference between $T_V$ and $A_V$, resulting in a larger theory prediction in comparison with the measured value.

The predicted branching fractions for  doubly Cabibbo-suppressed modes are suppressed by another order of magnitude with respect to those for  singly Cabibbo-suppressed ones because of the CKM matrix elements.  There are still many yet unobserved decays.  However, for those that have been observed, our predictions are consistent with the data within the 1-$\sigma$ range, except for the $D_s^+\to K^{*0}K^+$ decay whose measured value is significantly larger than theory predictions, though its error bar is also large.  The $D^+\to K^{*0}\pi^+$ and $D^+\to \rho^0 K^+$ modes involve respectively $A_V$ and $A_P$.  Without the contributions of $A_{P,V}$, their predicted branching fractions will be smaller than the measured values, clearly indicating the necessity of $A_{P,V}$.  In general, the predicted branching fractions of the doubly Cabibbo-suppressed modes under flavor SU(3) symmetry are more consistent with the data than the singly Cabibbo-suppressed modes.

For a comparison with our predictions, we have given ${\cal B}$(pole) and ${\cal B}$(FAT[mix]) in the last two columns of Tables~\ref{VPCF} to \ref{VPDCS}, transcribed from Ref.~\cite{Li:2013xsa} for the pole model~\cite{Li:2012cfa} and the factorization-assisted topological-amplitude (FAT) approach with the $\rho$-$\omega$ mixing, respectively.  The latter approach is preferred by the authors of Ref.~\cite{Li:2013xsa}.  Although ${\cal B}$(FAT[mix]) are generally in agreement with ours, there do exist some discrepancies.  For example, the predicted rates for both singly Cabibbo-suppressed $D^+\to\pi^+\omega$ and $D_s^+\to K^+\omega$ decays in the FAT[mix] approach are respectively much larger and smaller than ours.  As for the Cabibbo-allowed $D_s^+\to \rho^0\pi^+,\rho^+\pi^0$ modes, the FAT approach leads to vanishing rates for both of them \cite{Li:2013xsa}, while it is not so in our case. To see this, we notice that the topological amplitude expressions of $D_s^+\to \pi^+\rho^0$ and $D_s^+\to\pi^+\omega$ are given by
\begin{align}
\begin{split}
A(D_s^+\to \pi^+\rho^0) &= {1\over\sqrt{2}}V_{cs}^*V_{ud}(A_V-A_P) ~, \\
A(D_s^+\to \pi^+\omega) &= {1\over\sqrt{2}}V_{cs}^*V_{ud}(A_V+A_P) ~.
\end{split}
\end{align}
Moreover, we decompose the annihilation amplitude into
\begin{eqnarray}
A_{P,V}=a_{P,V}+A_{P,V}^r+A_{P,V}^e ~,
\end{eqnarray}
where $a$ is the short-distance $W$-annihilation amplitude, $A^r$ denotes the amplitude arising from resonant final-state interactions and the superscript $e$ indicates final-state rescattering via quark exchange. As shown in Ref.~\cite{Cheng:2010ry}, the $G$-parity argument implies that $a_V=-a_P$. Furthermore, the $D_s^+\to \pi^+\omega$ decay does not receive any resonant contribution, while rescattering via quark exchange is prohibited to contribute to $D_s^+\to \pi^+\rho^0$. Applying the relation \cite{FSI}
\begin{equation}
A_{P,V}^r=\frac{1}{2}(e^{2i\delta_r}-1)\left[a_{P,V}-a_{V,P}+\frac{1}{3}(C_{P,V}-C_{V,P})\right]
\end{equation}
for the nearby resonant contributions to $A_{P,V}$ induced by $C_{P,V}$ and
\begin{equation}
e^{2i\delta_r}=1-i\frac{\Gamma_R}{m_D-m_R+i\Gamma_R/2} ~,
\label{expRES}
\end{equation}
with $m_R$ being the resonance mass and $\Gamma_R$ its total decay width, we obtain
\begin{eqnarray}
A_V-A_P &=& 2a_V+(e^{2i\delta_r}-1)
\left[2a_V+{1\over3}(C_V-C_P)\right] ~,  \nonumber \\
A_V+A_P &=& A_V^e+A_P^e ~.
\label{eq:Arelation}
\end{eqnarray}
Therefore, while $D_s^+\to\pi^+\rho^0$ receives both short-distance and resonance-induced $W$-annihilation contributions, $D_s^+\to \pi^+\omega$ proceeds through long-distance final-state rescattering effects~\cite{Fajfer:2003ag}.  Hence, even if the short-distance annihilation amplitude is negligible, the former mode generally does not vanish in our consideration.  The small branching fraction $0.004\%$ quoted in Table~\ref{VPCF} for $D_s^+\to\pi^+\rho^0$ from the FAT[mix] approach comes from the $D_s^+\to \pi^+\omega$ decay followed by the $\rho$-$\omega$ mixing.

In addition to the decay $D_s^+\to\rho^+\eta'$ as discussed in passing,
Tables~\ref{VPCF} to \ref{VPDCS} also show that some experimental measurements are probably overestimated in the central values when compared with theory predictions: such as $D_s^+\to \overline{K}^0K^{*+}$, $D^0\to \overline K^0K^{*0}$, $D^+\to \overline K^0K^{*+}$ and $D_s^+\to K^+K^{*0}$.  The first mode was measured two decades ago \cite{CLEO1989}, and it is likely that the quoted experimental result for $D_s^+\to \overline K^{0}K^{*+}$ was overestimated. The predicted rates for $D^0\to \overline K^0K^{*0}$ and $D^0\to K^0\overline K^{*0}$ are the same, while the current limit is slightly below the prediction for the former.  We should stress that even though the central values of the current data for these modes may well be too large, the uncertainties associated with some of them are still quite big and await more precise measurements.

\section{SU(3) breaking effect}\label{sec:SU(3)breaking}

Supposing that the color-allowed and color-suppressed amplitudes are factorizable, they read
\begin{equation}
\begin{split}
\tilde T_V &=\frac{G_F}{\sqrt2}a_1(\ol K^*\pi)2f_\pi m_DA_0^{DK^*}(m_\pi^2) ~,\\
\tilde C_P &=\frac{G_F}{\sqrt2}a_2(\ol K^*\pi)2f_{K^*} m_DF_1^{D\pi}(m_{K^*}^2) ~,\\
\tilde T_P &=\frac{G_F}{\sqrt2}a_1(\ol K\rho)2f_\rho m_DF_1^{DK}(m_\rho^2) ~,\\
\tilde C_V &=\frac{G_F}{\sqrt2}a_1(\ol K\rho)2f_K m_DA_0^{D\rho}(m_K^2) ~,
\end{split}
\label{factorizationA}
\end{equation}
in the convention of Eq.~(\ref{decaywidthA}), and
\begin{equation}
\begin{split}
T_V &=\frac{G_F}{\sqrt2}a_1(\ol K^*\pi)2f_\pi m_{K^*}A_0^{DK^*}(m_\pi^2)(\epsilon\cdot p_D) ~,\\
C_P &=\frac{G_F}{\sqrt2}a_2(\ol K^*\pi)2f_{K^*} m_{K^*}F_1^{D\pi}(m_{K^*}^2)(\epsilon\cdot p_D) ~,\\
T_P &=\frac{G_F}{\sqrt2}a_1(\ol K\rho)2f_\rho m_\rho F_1^{DK}(m_\rho^2)(\epsilon\cdot p_D) ~,\\
C_V &=\frac{G_F}{\sqrt2}a_1(\ol K\rho)2f_K m_\rho A_0^{D\rho}(m_K^2)(\epsilon\cdot p_D)
\end{split}
\label{factorizationS}
\end{equation}
in the convention of Eq.~(\ref{decaywidthS}).
The decay constants to be used are $f_\pi=130.41$ MeV, $f_K=156.2$ MeV~\cite{pdg}, $f_{K^*}=220$ MeV and $f_\rho=216$ MeV~\cite{fV}.  We follow the definition of form factors in Ref.~\cite{BSW} and use the following parametrization~\cite{formfactor}
\begin{equation}
F(q^2)=\frac{F(0)}{(1-q^2/m_*^2)(1-\alpha q^2/m_*^2)} ~,
\end{equation}
where $m_*=m_{D_s^*}$, $m_{D_s}$, $m_{D^*}$ and $m_D$ when the form factors are $F_{1,0}^{DK}$, $A_0^{DK^*}$, $F_{1,0}^{D\pi}$ and $A_0^{D\rho}$, respectively. Form factors at $q^2=0$ and the parameter $\alpha$ are listed in Table~\ref{formfactor} (see \cite{Cheng:2010ry} for detail).  With the magnitudes and strong phases of $T_{P,V}$ and $C_{P,V}$ obtained in Section~\ref{fitting}, the Wilson coefficients $a_{1,2}$ can be extracted via Eqs.~(\ref{factorizationA}) and (\ref{factorizationS}).  The extracted $|a_{1,2}|$, $|a_2/a_1|$ and arg$(a_2/a_1)$ are listed in Table~\ref{a1a2} for different solutions.

\begin{table}[tp!]
\caption{Form factors at $q^2=0$ and the corresponding shape parameter $\alpha$.}
\begin{ruledtabular}
\begin{tabular}{l l l c c c c c c c c  c}
               &$F_0^{D\pi}$ &$F_0^{DK}$ &$F_1^{D\pi}$ &$F_1^{DK}$ &$A_0^{D\rho}$ &$A_0^{DK^*}$ \\
\hline
F(0)        &0.666            &0.739           &0.666             &0.739           &0.74                  &0.78\\
$\alpha$ &0.21              &0.30             &0.24               &0.33             &0.36                  &0.24\\
\end{tabular}
\label{formfactor}
\end{ruledtabular}
\end{table}

\begingroup
\squeezetable
\begin{table}[tp!]
\caption{The effective Wilson coefficients $a_{1,2}$, $|a_2/a_1|$ and arg$(a_2/a_1)$ extracted from the Cabibbo-favored $D^+\to\ol{K}^{*0}\,\pi^+$ and $\ol{K}^0\,\rho^+$ decay modes based on Solutions (A1), (A5), (S4) and (S5) shown in Table~\ref{globalA} and Table~\ref{globalS}.}
\begin{ruledtabular}
\begin{tabular}{l l l c c c c c c c c}
 & \multicolumn{4}{c}{$\ol{K}^{*0}\,\pi^+$} & \multicolumn{4}{c}{$\ol{K}^0\,\rho^+$} \\
\cline{2-5} \cline{6-9}
                        &(A1)                   &(A5)                 &(S4)                  &(S5)                    &(A1)                   &(A5)               &(S4)  &(S5)\\
\hline
$|a_1|$               &$1.34\pm0.06$ &$1.22\pm0.05$&$1.45\pm0.07$ &$1.20\pm0.07$&$1.43\pm0.04$  &$1.43\pm0.04$&$1.38\pm0.11$&$1.43\pm0.04$\\
$|a_2|$               &$0.69\pm0.06$ &$0.85\pm0.02$&$0.73\pm0.06$ &$0.80\pm0.02$&$1.05\pm0.05$  &$1.04\pm0.06$&$1.09\pm0.03$&$1.07\pm0.05$\\
$|a_2/a_1|$      &$0.52\pm0.05$ &$0.69\pm0.03$&$0.50\pm0.05$ &$0.66\pm0.04$&$0.73\pm0.04$  &$0.73\pm0.04$&$0.79\pm0.07$&$0.75\pm0.04$\\
arg$(a_2/a_1)$&$-(157\pm2)^\circ$ &$-(165\pm3)^\circ$&$-(159\pm2)^\circ$ &$-(162\pm3)^\circ$&$(158\pm51)^\circ$  &$(158\pm36)^\circ$&$-(151\pm141)^\circ$&$(157\pm42)^\circ$\\
\end{tabular}
\label{a1a2}
\end{ruledtabular}
\end{table}
\endgroup

If we assume for factorizable amplitudes that the effective Wilson coefficients $a_{1,2}$ are the same.  Then their magnitudes will differ mode by mode due to differences in the final-state meson masses, decay constants, and form factors.  For the singly Cabibbo-suppressed decay modes, the predicted ${\cal B}(D^+\to\pi^+\phi)$ in Solution (A1) has the largest deviation from the current data.  Its factorizable amplitude is
\begin{equation}
C'_{P,\pi^+\phi}=\frac{G_F}{\sqrt2}a_22f_\phi m_{D^+}F_1^{D\pi}(m_\phi^2) ~.
\end{equation}
Comparing with the related Cabibbo-favored $D^+\to\ol{K}^{*0}\,\pi^+$ decay mode, we obtain the ratio
\begin{equation}
\frac{C'_{P,\pi^+\phi}}{ C_{P,\ol{K}^{*0}\,\pi^+}}=\frac{f_\phi}{f_{K*}}\frac{F^{D\pi}_1(m_\phi^2)}{F^{D\pi}_1(m_{K^{*0}}^2)} \simeq 1.07 ~,
\end{equation}
where $a_2$ for these two decay modes is assumed to be the same and cancels out.  Including this symmetry breaking factor, the invariant decay amplitude of $D^+\to\pi^+\phi$ now becomes
\begin{equation}
{\cal A}(D^+\to\pi^+\phi)=1.07\times Y_sC_P' ~.
\label{symmbreakKK}
\end{equation}
As a consequence, the predicted branching fraction is enhanced from $(3.92\pm0.69)\times10^{-3}$ to $(4.49\pm0.80)\times10^{-3}$, closer to the current data of $(5.66^{+0.19}_{-0.21})\times10^{-3}$.  The predicted ${\cal B}(D^+\to\pi^+\phi)$ in Solution (S4) also deviates from the measurement most significantly among all the singly Cabibbo-suppressed modes.  By the same token, our prediction is enhanced from $(3.37\pm0.59)\times10^{-3}$ to $(4.50\pm0.87)\times10^{-3}$ after taking the symmetry breaking effect into account, but using Eq.~(\ref{factorizationS}) in this case. This method is also applicable to the $D^0\to\pi^0\phi$ decay.

Even though the uncertainty associated with the current data of ${\cal B}(D^+\to\ol K^0K^{*+})$ is still quite large, the central value of our prediction for this mode is more than two times smaller, and so are the other predictions made in the pole model and the FAT approach with the $\rho$-$\omega$ mixing.  The factorizable amplitude for this mode is
\begin{align}
T'_{P,\ol K^0K^{*+}}
= \frac{G_F}{\sqrt2}a_12f_{K^*}m_{D}F_1^{DK}(m_{K^{*+}}^2)~.
\end{align}
Comparing with the $D^+\to \ol K^0\rho^+$ decay, we obtain the ratio
\begin{align}
\frac{T'_{P,\ol K^0K^{*+}}}{T_{P,\ol K^0\rho^+}}=\frac{f_{K^*}}{f_\rho}\frac{F^{DK}_1(m_{K^{*+}}^2)}
{F^{DK}_1(m_{\rho^+}^2)}\simeq 1.09 ~.
\end{align}
Therefore, the flavor amplitude of $D^+\to\ol K^0K^{*+}$ now becomes
\begin{align}
{\cal A}(D^+\to \ol K^0K^{*+})=Y_dA_P'+1.09\times Y_sT_P'~.
\end{align}
The predicted ${\cal B}(D^+\to\ol K^0 K^{*+})=(14.45\pm2.45)\times10^{-3}$ in Solution (A1) is thus enhanced to $(17.10\pm2.69)\times10^{-3}$ whose central value now becomes slightly closer to the current data.

Although some of the modes will have better agreement with the data after the above-mentioned symmetry breaking is included, some others will deviate from the measurement even more regardless which solution we take. Take the decay $D^+\to K^+\ol K^{*0}$ as an example.  Its factorizable amplitude $T'_V$ is written as
\begin{align}
T'_{V,K^+\ol K^{*0}}
= \frac{G_F}{\sqrt2}a_12f_Km_{D}A_0^{DK^{*}}(m_{K^+}^2)~.
\end{align}
Comparing with the factorization amplitude of the mode $D^+\to\ol K^{*0}\pi^+$, we have
\begin{align}
\frac{T'_{V,K^+\ol K^{*0}}}{T_{V,\ol K^0\pi^+}}=\frac{f_K}{f_\pi}\frac{A^{DK^*}_0(m_{K^{+}}^2)}{A^{DK^*}_0(m_{\pi^+}^2)}\simeq 1.28~.
\end{align}
Hence, the flavor amplitude of the this mode becomes
\begin{align}
{\cal A}(D^+\to K^+\ol K^{*0})=Y_dA_V'+1.28\times Y_sT_V'~,
\end{align}
and the predicted branching fraction is enhanced.  Using Solution (A1), the predicted branching fraction of $(4.00\pm0.82)\times10^{-3}$ based on exact flavor SU(3) symmetry now becomes $(6.4\pm1.1)\times10^{-3}$ which deviates even more from the current data $(3.84^{+0.14}_{-0.23})\times10^{-3}$.

We also list the results for Solutions (A5) and (S5) in Table~\ref{a1a2}.  Although both of them are disfavored by many of the singly Cabibbo-suppressed decay modes, their extracted $|a_2/a_1|$ for different decay modes are much closer to each other.  In spite of the fact that taking into account the symmetry breaking factors in the factorizable amplitudes results in more deviation from the experimental data for modes like $D^+\to K^+\ol K^{*0}$, such factors in other singly Cabibbo-suppressed modes do improve agreement, as illustrated above in the two examples of $D^+\to\pi^+\phi$ and $\ol K^0K^{*+}$.  Nevertheless, it is pertinent to conclude that the flavor SU(3) symmetry is generally a good approximate symmetry in explaining the $D\to V\!P$ data.

\section{Conclusions}\label{sec:conclusions}

Because of the low masses of charmed mesons, their hadronic decays are best analyzed using the diagrammatic approach with the assumption of flavor SU(3) symmetry.  Within this framework and using the latest data, we have updated the global $\chi^2$ fit to the Cabibbo-favored decay branching fractions and, thanks to the recent measurement of ${\cal B}(D_s^+\to\pi^+\rho^0)$, determined for the first time the $W$-annihilation amplitudes $A_{P,V}$.  They are the smallest in size among all the tree-level flavor amplitudes analyzed in this work.  A determination of ${\cal B}(D_s^+\to\pi^0\rho^+)$ will be very useful in confirming the information we get from ${\cal B}(D_s^+\to\pi^+\rho^0)$ and reducing the uncertainties associated with $A_{P,V}$.  During the fits, we have found several possible solutions.  Many of them are ruled out by the the data of singly Cabibbo-suppressed modes.

Using the flavor amplitudes extracted from the Cabibbo-favored decays, we are able to predict the branching fractions of all the $D\to V\!P$ decays under flavor SU(3) symmetry and testing this working assumption, particularly in the Cabibbo-suppressed decays.
The predictions for the doubly Cabibbo-suppressed channels are in good agreement with the data, while some of those for the singly Cabibbo-suppressed decay modes are seen to violate the flavor SU(3) symmetry.  We have tried to include SU(3) symmetry breaking in color-allowed and color-suppressed tree amplitudes to see if a better agreement with data can be achieved.  However, the conclusion is mixed, and the exact flavor SU(3)-symmetric approach is still sufficiently adequate to provide an overall explanation for the current data.

We have also compared our diagrammatic-approach results in some detail to those of other existing theoretical calculations in the literature.  In order to test which theories are more favored by Nature, we need to await more precisely measured data, especially those of yet unobserved modes and some of the singly Cabibbo-suppressed decays that have significant deviations from theory predictions.


\section{Acknowledgments}

This work was supported in part by the Ministry of Science and Technology of Taiwan under Grant Nos.~MOST~104-2112-M-001-022 and MOST~104-2628-M-008-004-MY4.


\end{document}